\documentclass[useAMS,usenatbib]{mn2e}
\usepackage{lmodern}
\usepackage[utf8]{inputenc}
\usepackage[english]{babel}
\usepackage[usenames,dvipsnames,svgnames,table]{xcolor}
\usepackage{ulem}
\usepackage{amsmath}
\usepackage{graphicx}
\usepackage{subfig}

\newcommand{\guy}[1]{#1}
\newcommand{\gr}[1]{#1}

\newcommand{\Kepler}[1]{\textit{Kepler}#1}
\newcommand{\Corot}{CoRoT}

\title[Correcting reported pulsation frequencies]{Why should we correct reported pulsation frequencies for stellar line-of-sight Doppler velocity shifts?}
\author[G. R. Davies et al.]{G.R.~Davies$^{1,2}$\thanks{E-mail: davies@bison.ph.bham.ac.uk}, R.~Handberg$^{1,2}$, A.~Miglio$^{1,2}$, T.L.~Campante$^{1,2}$, \newauthor W.J.~Chaplin$^{1,2}$, Y.~Elsworth$^{1,2}$ \\
$^1$ School of Physics and Astronomy, University of Birmingham, Edgbaston, Birmingham B15 2TT, United Kingdom\\
$^2$ Stellar Astrophysics Centre (SAC), Department of Physics and Astronomy, Aarhus University, DK-8000 Aarhus C, Denmark}
\date{Received 2014 August 28 / Accepted 2014 August 29}
\begin{document}
\maketitle
\label{firstpage}
\begin{abstract}
In the age of \Kepler{ }and \Corot, extended observations have provided estimates of stellar pulsation frequencies that have achieved new levels of precision, regularly exceeding fractional levels of a few parts in $10^{4}$.  These high levels of precision now in principle exceed the point where one can ignore the Doppler shift of pulsation frequencies caused by the motion of a star relative to the observer.  We present a correction for these Doppler shifts and use previously published pulsation frequencies to demonstrate the significance of the effect.  We suggest that reported pulsation frequencies should be routinely corrected for stellar line-of-sight velocity Doppler shifts, or if a line-of-sight velocity estimate is not available, the frame of reference in which the frequencies are reported should be clearly stated.
\end{abstract}
\section{Introduction}\label{sec:intro}
It is a golden age for asteroseismology with both \Kepler{ }\citep{2010PASP..122..131G} and CoRoT \citep{2008Sci...322..558M,2008A&A...488..705A} having provided high-quality time series for analysis.  In the coming years, missions like K2, TESS, PLATO, and SONG will further our understanding of the Galaxy and our place in it, and asteroseismology will significantly contribute to this.  Asteroseismology has already allowed us to study a large number of stars in detail \citep[see][for a review on solar-like oscillations]{2013ARA&A..51..353C} based on measured frequencies of the rich spectra of overtones shown by these stars.\\
It is well known that measured frequencies and periods must be corrected for Doppler velocity shifts between the source and the observer.  \guy{\Kepler{ }time series are routinely corrected for spacecraft orbital motion equivalent barycentric arrival times, while \Corot{ } time series are corrected to the slightly less preferable heliocentric arrival times.}  Although it is known that a Doppler shift from the motion of a star relative to, for example, the solar system barycentre shifts measured frequencies \citep[e.g. see][]{1992A&A...253..178P, 1995A&A...295L..17P, 2005ASPC..335..115R}, this correction has been routinely ignored, most likely due to it being in the majority cases smaller than the quoted frequency uncertainties. However, recent observational developments in the field -- especially long timeseries data collected by \Corot{ }and \Kepler{,} notably on solar-like oscillators -- are providing us with frequencies of high precision which the community is starting to exploit for detailed stellar modelling. It is therefore crucial that accurate intrinsic stellar oscillation frequencies are reported.\\
In this paper we consider the impact of Doppler shifts due to stellar motion on previously published frequencies, for different categories of pulsating stars. We show that the correction is, in many cases, \guy{ statistically significant}.  Finally, we provide a brief discussion on the significance of this systematic effect on the estimates of fundamental stellar properties.
\section{The effect}\label{effect}
The relativistic Doppler effect for motion along the line of sight is well defined.  Consider the radial velocity $v_{\rm r}$, with $\beta = v_{\rm r}/c$, \guy{such that a positive velocity produces a redshift}.  If the frequency emitted by the source is $\nu_{\rm s}$ and the observed frequency is $\nu_{\rm o}$, then the two frequencies are related by \guy{, and furthermore well approximated if $v \ll c$:
\begin{equation}
\nu_{\rm s} = \sqrt{\frac{1+\beta}{1-\beta}} \, \nu_{\rm o} \simeq (1 + \beta) \nu_{\rm o}.
\label{eq::corr}
\end{equation} 
}
When spectroscopic observations provide an estimated line-of-sight velocity shift $v_{\rm spec}$ in the solar system barycentric frame, the Doppler shift is calculable.  By applying Equation \ref{eq::corr} to the observed frequencies, $\nu_{\rm o}$, we may obtain estimates of the frequencies $\nu_{\rm s}$ in the frame of reference of the target star. These are the Doppler-shift corrected frequencies that we seek.\\
Measured spectroscopic radial velocities contain contributions from effects that are not strictly speaking line-of-sight velocities.  For the Sun there are contributions from gravitational redshift ($v_{\rm grs, \odot} \approx 633 \, \rm m \, s^{-1} $) and convective blueshift (a few tens of $\rm m \, s^{-1}$).  For white dwarfs, the gravitational redshift can be of the order of a few tens of $\rm km \, s^{-1}$.  We will assume the effect of convective blueshift to be negligible.\\
\gr{Gravitational redshift is a consequence of gravitational time dilation, so that signals emitted in lower gravitational potentials are redshifted.  On the assumption that spectroscopic observables and pulsation signals are emitted at the same height in the stellar photosphere, both signals will be redshifted due to time dilation and both signals will be redshifted by the same amount.  Hence a correction for the line-of-sight Doppler shift that uses a spectroscopic measure which includes a contribution from $v_{\rm grs}$ will correct for the gravitational time dilation of the emitted pulsation signal.  Hence, we can estimate,\guy{
\begin{equation}
\beta = \frac{v}{c} \approx \frac{v_{\rm spec}}{c}.
\end{equation}  
}
We assume here that the observer, for both spectroscopic and pulsation observations, is at the same distance from the source and not in an independent gravitational potential well of their own.
For \guy{most} purposes it will be sufficient to treat uncertainties in Doppler shifted pulsations as, \guy{
\begin{equation}
\sigma_{\nu_{\rm s}} \approx \sqrt{\frac{1 + \beta}{1 - \beta}} \, \sigma_{\nu_{\rm o}} \simeq (1 + \beta) \, \sigma_{\nu_{\rm o}} .
\end{equation}}
}

\section{The precision and accuracy of pulsation frequencies}
\subsection{The distribution of $v_{\rm spec}$ in the \Kepler{ }field}
\begin{figure}
\centering
\subfloat[]{\includegraphics[width=88mm]{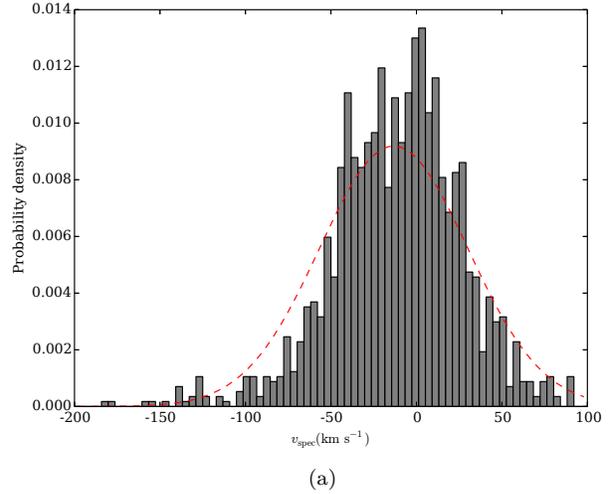}}\\%
\caption{The distribution of spectroscopic redshift velocities from APOGEE for a sample stars in the \Kepler{ }field of view.  The red curve shows a normal distribution with mean $-13 \rm \, km \, s^{-1}$ and standard deviation $43 \rm \, km \, s^{-1}$.}
\label{fig::APO}
\end{figure}

Full exploitation of the potential of asteroseismic observations requires complementary data from follow-up observations. Ground-based spectroscopic campaigns can provide the $T_{\rm eff}$ and $\rm [Fe/H]$ \citep[see e.g.][]{2004A&A...425..683B, 2012MNRAS.423..122B,2013MNRAS.434.1422M} required for determining fundamental stellar parameters such as mass, radius, and age.  A natural output of this spectroscopic work is the radial velocity of the star.  The precision of the determined radial velocity is typically much better than $1 \rm \, km \, s^{-1}$.  \guy{Typically, the uncertainty on the source frequency can be approximated as the observed frequency uncertainty multiplied by $1 + \beta$.}  In this limit, uncertainties on the stellar pulsation frequencies are not increased significantly by applying the correction.\\
Figure \ref{fig::APO} shows a distribution of measured radial velocities for a population of stars in the \Kepler{ }field taken from spectroscopic data collected by the SDSS-III \guy{Apache Point Observatory Galactic Evolution Experiment} \citep{2014ApJS..211...17A}.  We note that the median of the distribution is $-13 \rm \, km \, s^{-1}$ with a standard deviation of $43 \rm \, km \, s^{-1}$.
The order of magnitude for a typical Doppler shift correction is $v/c \approx -13 \rm \, km \, s^{-1} / c \approx 10 ^{-4}$ which might initially be regarded as a small effect.  However, the quoted uncertainties on pulsation frequencies measured for coherent oscillations or more recently by \Kepler{ }and CoRoT for stochastic oscillations are regularly below this threshold.  
\subsection{Example stars}
\begin{figure*}
\centering
\subfloat[]{\includegraphics[width=160mm, height=90mm]{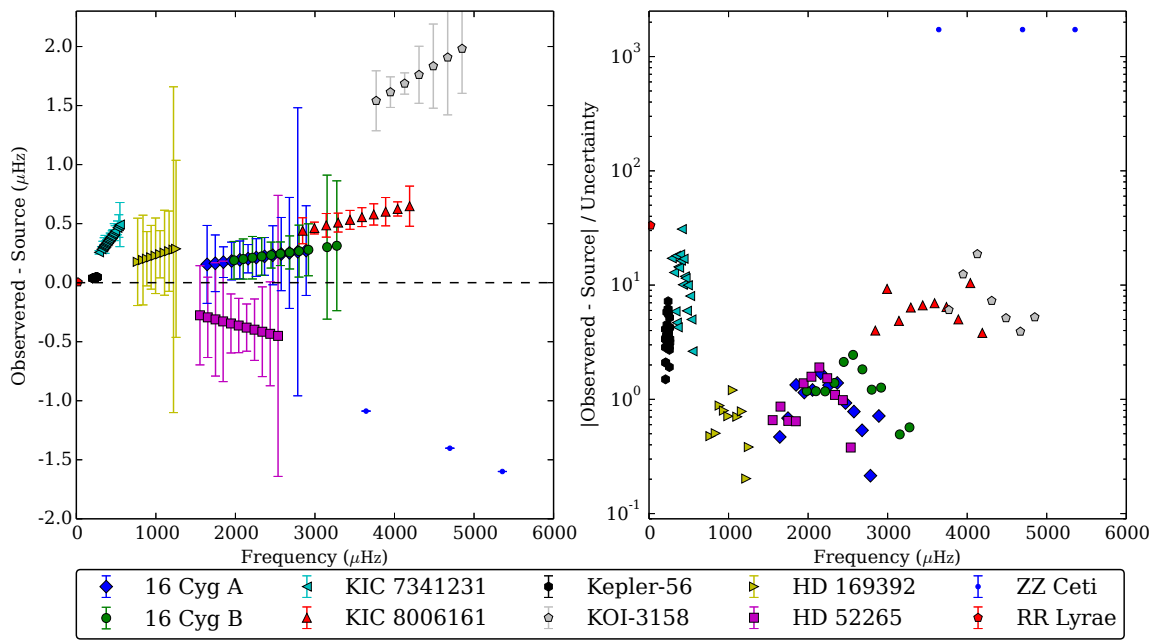}}%
\caption{Differences in published (e.g. solar-system barycentric frame of reference) and Doppler shift corrected (source frame of reference) pulsation frequencies for a selection of stars and modes of oscillation.  The left panel gives the differences as a function of frequency while the right panel displays the difference divided by the uncertainty.}
\label{fig::shifts2}
\end{figure*}

To demonstrate that the Doppler shift correction is significant in terms of precision we have selected a number of stars of interest with published pulsation frequencies.  Table \ref{tab::pubs} lists the example stars together with the parameters required to apply the Doppler shift correction.  \\
Our selection of stars includes a Galactic halo member (KIC 7341231) and at least one thick disk member (KOI-3158).  In terms of stellar pulsation classification, our list is dominated by solar-like oscillators but does include two classical pulsators that define their classes, namely ZZ Ceti and RR Lyrae.  Because of its high amplitude pulsation, RR Lyrae undergoes significant changes to its radius and here we adopt a radius consistent with a minimum value.  In addition, \cite{1999ApJS..121....1M} do not supply an uncertainty for the spectroscopic radial velocity so we have adopted an uncertainty of $0.1 \rm \, km \,s^{-1}$.\\  
The precision in the measurement of the spectroscopic redshift is high for all the listed stars, which means that the uncertainties in the pulsation frequencies in the frame of the star will not be significantly increased with respect to the uncertainties of the uncorrected frequencies.\\
Figure \ref{fig::shifts2} shows the difference between corrected and uncorrected frequencies and then the same differences divided by the frequency uncertainties.  Frequencies shown are a selection of the published frequencies.  For the solar-like oscillators, only the dipole modes of oscillation are shown and modes with large error bars (and hence little constraint) are not plotted to avoid overcomplicating the plot.  It is clear that corrections for the Doppler velocity shift can be highly significant.\\  
The published frequencies of oscillation were extracted from datasets that are shorter than those available from the longest \Kepler{ }time series ($\approx 1470$ days).  As detailed analyses are performed with the full data sets, the precision in the frequencies is expected to increase. As an example, the published frequencies of the solar analogues 16 Cyg A and B come from 3 months of data.  If the full compliment of \Kepler{ }data is used (around 30 months) the precision can be expected to improve as $T^{-1/2}$ where $T$ is the length of the data set \guy{\citep{1992ApJ...387..712L}}.  Hence, the 16 Cyg frequencies from the full \Kepler{ }data set may be as much as 3 times more precise than the current published values.  This increase in precision makes a correction for Doppler shifts even more significant.
\begin{table*}
\centering 
  \caption{Example stars and adopted properties.  For the solar-like oscillators 16 Cyg A \& B, HD173701, KOI-3158, KIC 7341231, and Kepler-56 we have used results from observations taken by \Kepler{.  }Frequencies for the solar-like oscillators HD169392 and HD52265 have been taken from CoRoT observations.  \Kepler{ }has observed the classical pulsator RR Lyrae (for RR Lyrae we ignore the variation of the pulsation frequency caused by the Blazhko effect and also ignore the line-of-sight velocity due to the pulsation itself that may contribute of order upto ten $\rm km \, s^{-1}$) and the white dwarf ZZ Ceti has been observed as part of a ground based photometric campaign.}
  \begin{tabular}{cccccc}
  \hline
  ID & Mass & Radius  & Typical frequency & $v_{\rm spec}$ & Source of Published \\
  	& ($M_{\odot}$) & ($R_{\odot}$)  & ($\rm \mu Hz$) & ($\rm km \, s^{-1}$) & frequencies \\
 \hline
  16 Cyg A & $1.11 \pm 0.02^{\rm a}$ & $1.243 \pm 0.008^{\rm a}$ & $2201$ & $-27.61 \pm 0.08^{\rm d}$ & \cite{2012ApJ...748L..10M} \\
  16 Cyg B & $1.07 \pm 0.02^{\rm a}$ & $1.127 \pm 0.007^{\rm a}$ & $2567$ & $-28.02 \pm 0.08^{\rm d}$ & \cite{2012ApJ...748L..10M} \\ 
  HD 173701 & $1.114 \pm 0.017^{\rm f}$ & $0.99 \pm 0.015^{\rm f}$ & $3550$ & $-45.64 \pm 0.08^{\rm d}$ &  \protect{\cite{2012A&A...543A..54A}} \\
  KOI-3158 & $0.758 \pm 0.043^{\rm p}$ & $0.752 \pm 0.014^{\rm p}$ & $4500$ & $-121.9 \pm 0.1^{\rm h}$ & Campante et al. (Sub) \\
  KIC 7341231 & $0.77 - 0.88^{\rm b}$ & $2.6^{b}$ & $406$ & $-269.16 \pm 0.14^{\rm e}$ &  \cite{2012ApJ...756...19D} \\
  Kepler-56 & $1.32 \pm 0.13^{\rm c}$ & $4.23 \pm 0.15^{\rm c}$ & $244$ & $-54.4 \pm 0.1^{\rm x}$ &  \cite{2013Sci...342..331H} \\  
  HD169392 &  $1.15 \pm 0.01^{\rm j}$ & $1.88 \pm 0.02^{\rm j}$ & $1030$ & $-68.1606\pm 0.0004^{\rm j}$ &  \protect{\cite{2013A&A...549A..12M}} \\
  HD52265 & $1.24 \pm 0.02^{\rm k}$ & $1.33 \pm 0.02^{\rm k}$ & $2000$ & $53.86 \pm 0.09^{\rm d}$ &  \protect{\cite{2011A&A...530A..97B}} \\
  ZZ Ceti & $0.609 \pm 0.012^{\rm l}$ & $0.149 \pm 0.001^{\rm l}$ & $4000$ & $91.2^{\rm m}$ & \cite{2013ApJ...771...17M} \\
  RR Lyrae & $0.65^{\rm n}$ & $5.1^{\rm adopted}$ & $20$ & $-56.48 \pm 0.06^{\rm o}$ & \cite{2011MNRAS.411..878K} \\
\hline
\end{tabular}\\
$^{\rm a}$ \citep{2012ApJ...748L..10M}, $^{\rm b}$ \citep{2012ApJ...756...19D}, 
$^{\rm c}$ \citep{2013Sci...342..331H}, $^{\rm d}$ \citep{2002ApJS..141..503N}, \\
$^{\rm e}$ \citep{2002AJ....124.1144L}, $^{\rm f}$ \citep{2013MNRAS.435..242G}, 
$^{\rm g}$ \citep{2014ApJS..210....1C}, $^{\rm h}$ \citep{2004A&A...418..989N}\\
$^{\rm j}$ \citep{2013A&A...549A..12M}, $^{\rm k}$ \citep{2012A&A...543A..96E}
$^{\rm l}$ \citep{2013ApJ...771...17M}, $^{\rm m}$ \citep{1999ApJS..121....1M} \\
$^{\rm n}$ \citep{2010A&A...519A..64K}, $^{\rm o}$ \citep{2013ApJ...773..181N}, \\
$^{\rm p}$ Campante et al. (Submitted), $^{\rm x}$ Buchhave private comm.
\label{tab::pubs}
\end{table*}
\section{Discussion}
The impact of this Doppler correction on reported fundamental properties and parameters is difficult to assess.  For a single star, the correction we detail here is a straightforward correction for a bias that is systematic on all pulsation frequencies.  One might reasonably argue that the correction should be applied no matter what the perceived impact on measured fundamental properties.  However, that said there are challenges in comparing observed pulsation frequencies with model frequencies beyond this relatively small effect.  For example, there is currently significant uncertainty in the physics of modelled stars and their modelled pulsation frequencies, in many cases a dominant component of any differences.
\subsection{Solar-like pulsations}
Solar-like pulsations are characterised by modes that are stochastically excited and intrinsically damped by near-surface convection.  Modes have finite lifetimes and hence resonant peaks in the frequency domain typically have width that impacts the uncertainty in the pulsation frequencies.  Precision in solar-like modes of pulsation is much less than that of classical pulsations and hence the level of effect for an individual mode is small but nevertheless often (just about) significant.  However, many tens of modes may be detected in a solar-like oscillator, and the Doppler shift effect does not cancel out from mode to mode.  This can make the shift very significant in terms of a simple observed-to-corrected $\chi^{2}$ statistic.\\
For more evolved sub-giant and red giant stars, modes of mixed character (mixed acoustic and buoyancy modes) are observable with very long life times, implying narrow width in the frequency domain and high levels of precision.  Here the size of the frequency shift is tempered by more evolved stars having observable modes of oscillation that lie at lower frequencies.  However, the example red giant star Kepler-56 shows frequency shifts of very high significance.\\
The largest impact on fundamental properties comes when observed individual frequencies have been directly compared to modelled frequencies.  However, much of the effect will be mitigated by the treatment of the so called ``surface correction'' .  In fact, a Doppler shift applied to all frequencies by a constant factor is equivalent to a homologous scaling \cite[see the factor $r$ in][]{2008ApJ...683L.175K}.  Hence, the surface term has the ability to absorb, to some degree, systematics introduced by Doppler shifts.\\ 
It is also possible to model stars by using combinations of frequencies, frequency differences or ratios, in order to mitigate the effects of the surface term and this will suppress significantly the impact of any Doppler shift.  Because of the linear nature of the Doppler correction, differences in frequency are fractionally shifted by approximately $\beta $.  If the fractional uncertainty in the difference in frequency is much larger than $\beta$, then the significance of the impact of the Doppler effect is negligible.  Hence, for stars with parameters and properties that are well characterised by frequency spacings (e.g. rotational splitting), this Doppler correction is unlikely to be significant.  Even more so, for analyses that use frequency ratios (that is ratios of similar frequencies; \citealt{2003A&A...411..215R}) the Doppler correction is very, very small.\\
Some stars might also be expected to show frequency shifts due to stellar activity cycles.  In the Sun observed as a star, pulsation frequencies are shifted by as much as $0.5 \rm \, \mu Hz$ which is equivalent to a Doppler velocity shift of around $50 \rm \, km \, s^{-1}$.  However, follow-up observations, e.g. monitoring radial velocity, star-spot activity, or calcium II H \& K lines, have the potential to distinguish between a changing radial velocity and genuine activity-cycle shifts.
\subsection{Classical pulsators}
Here we characterise classical pulsators as stars with coherent or near coherent, pulsations.  The frequencies of observed coherent pulsations can be estimated to a very high level of precision.  This means that even quite small Doppler shifts can give rise to significant corrections to the observed frequencies.  However, small effects may not be important to fundamental properties. Modelled frequencies of pulsation for many classes of classical pulsator cannot achieve the required precision to be fully exploit the observed precision.  This may be a limiting factor when considering the impact on fundamental properties.\\
With the increased precision for classical pulsators comes the ability to detect and identify small effects, for example signals from orbital companions or secular stellar  evolution.\\  
So far we have assumed that there is no change in the observer-source line-of-sight velocity.  It is possible to detect changes in pulsation frequency due to a secular change in radial velocity \citep{1995A&A...295L..17P}.  Significant changes that occur on time scales of the same order or less than the length of observation should be treated differently to the above.  If the temporal evolution of the radial velocity is well known and precise then a correction can be applied directly to the observation time stamps.  If the change in line-of-sight velocity is not known it is possible to estimate changes from the evolution of the pulsations \citep{1992A&A...253..178P,2012MNRAS.422..738S,2014arXiv1404.5649M}.  Multi-star or star-planet systems are candidates for this type of analysis.  We also note that line-of-sight Doppler shifts will be present in measured planetary and eclipsing binary periods, often quoted with fractional precisions of better than $10^{-5}$.
\section{Conclusion} 
Methods for estimating the frequencies of stellar pulsations should ideally include an effort to account for all forms of uncertainty, at least where a correction is significant and warranted.  Here we show the trivial method to account for line-of-sight Doppler velocity shifts.  To apply such a correction, an estimate of the line-of-sight velocity is required.  We suggest that spectroscopic studies of pulsating stars routinely publish the measured radial velocity in the solar system barycentric frame of reference for use in asteroseismic analysis.  \\
We believe it to be a matter of principle that published frequencies and their uncertainties should account for any known, tractable, and non-negligible bias (while also stating limitations that cannot be overcome).  In the case of a known Doppler shift, the effect is understood, the correction is trivial, and we have shown the effect to be significant on individual frequencies for a number of stars.  
We would propose that future asteroseismic analyses \guy{publish frequencies corrected for Doppler shifts, together with the adopted spectroscopic radial velocity and the date of observation.  This will prevent ambiguity and allow for the possibility of revision if more suitable spectroscopic observations become available.  In the absence of a $v_{\rm spec}$ measurement, the uncorrected frequencies should be published and labelled clearly.  In terms of stellar modelling, one could use a prior probability distribution, formed from measurements similar to Figure \ref{fig::APO}, to fully account for the additional uncertainty.}\\
While at present it may be true that model predicted pulsation frequencies cannot make full use of the the precision provided by observations, one would hope that that this will not remain the case in the coming years.  Hence we believe that published frequencies should make this Doppler correction.  Alternatively, sufficient information should be provided to allow such a correction, e.g. one should state which frame of reference the published frequencies are expressed in\footnote{At the time of publication, the standard pipeline for \Kepler{ }observations correct to the solar system barycentric frame, while CoRoT time series are corrected to the heliocentric frame.}.
\section{Acknowledgements}
The authors with to thank Tim Bedding, Fr\'{e}d\'{e}ric Baudin, Lars Buchhave, \guy{Clive Speake}, Daniel Huber, Steve Kawaler, Ennio Poretti, V\'{i}ctor Silva Aguirre, and Roberto Silvotti.  The authors acknowledge the support of the UK Science and Technology Facilities Council (STFC).  Funding for the Stellar Astrophysics Centre is provided by The Danish National Research Foundation (Grant agreement no.: DNRF106). The research is supported by the ASTERISK project (ASTERoseismic Investigations with SONG and \Kepler{) }funded by the European Research Council (Grant agreement no.: 267864).
\bibliographystyle{mn2e_new}
\bibliography{refs}
\label{lastpage}
\end{document}